\documentclass{ws-procs9x6}
\def\poinc{Poincar\'{e} }
\def\bfq {{\bf q}}
\def\bfK{{\bf K}}
\def\bfk{{\bf k}}
\def\bfp{{\bf p}}  
\def\be{\begin{equation}}
 \def \ee{\end{equation}}
\def\bea{\begin{eqnarray}}
  \def\eea{\end{eqnarray}}

\def\notp{{\not\! p}}
\def\notk{{\not\! k}}
\def\up{{\uparrow}}
\def\down{{\downarrow}}

\begin{document}

\title{The Electromagnetic  Form Factors
              of the Proton and Neutron: Fundamental
              Indicators of Nucleon Structure}

\author{Gerald A. Miller}

\address{Department of Physics, University of Washington\\
  Seattle, WA 98195-1560\\
E-mail: miller@phys.washington.edu}


\maketitle

\abstracts{We present a relativistic
  interpretation for why the proton's
  $G_E/G_M$ falls and 
$QF_2/F_1$ is approximately constant. 
  Reproducing 
the observed  $G_E^n$ then mandates  the inclusion of the
effects of the pion cloud.}

\section{Introduction}

An alternate title would be ``Surprises in the Proton''. This talk owes its
existence to the brilliant, precise, stunning and exciting recent
experimental work on measuring $G_E/G_M$ (or $QF_2/F_1$) for the proton and
$G_E$ for the neutron. My goal here is to interpret the data. Symmetries
including
Poincar\'{e} invariance and chiral symmetry will be the principal tool
I'll use.

If, a few years ago,
one had asked participants at a meeting like this about the $Q^2$ dependence
of the proton's $G_E/G_M$ or $QF_2/F_1$. Almost everyone one have answered that
for large enough values of $Q^2$, $G_E/G_M$ would be flat and
$QF_2/F_1$ would fall with increasing  $Q^2$. The reason for the latter fall
being conservation of hadron helicity. 
Indeed, the shapes of the curves have been
obtained in the new measurements, except for the mis-labeling of the ordinate
axes. The expected flatness of $G_E/G_M$ holds for $QF_2/F_1$, and the
 quantity $G_E/G_M$ falls rapidly and linearly with $Q^2$. This revolutionary
 behavior needs to be understood!

\section{Outline}

I will begin with a brief discussion of Light Front Physics. Stan Brodsky has
long been advocating this technique, I have become a convert.
Then I will discuss a particular relativistic model of the nucleon,
and  proceed to apply it. The proton calculations will be discussed first,
but recent high accuracy experiments make it necessary for us to compute
observables for the neutron as well.               

\section{Light Front }

Light-front dynamics
is a relativistic many-body dynamics in which fields are quantized at a
``time''=$\tau=x^0+x^3\equiv x^+$. The $\tau$-development operator
 is then given by
$P^0-P^3\equiv P^-$. These equations show the  notation that    
a four-vector $A^\mu$ is expressed  as
$ A^\pm\equiv A^0\pm A^3.$
One quantizes at $x^+=0$ which is a light-front, hence the name ``light front
dynamics''.
 The canonical spatial variable must be orthogonal to the time variable,
 and this is given by 
$x^-=x^0-x^3$. The canonical momentum is then $P^+=P^0+P^3$. The other
coordinates are  ${\bf x}_\perp$ and  ${\bf P}_\perp$.

 The most important  consequence of this is that the
 relation between energy and
momentum of a free particle is given by:
$ p_\mu p^\mu=m^2=p^+p^--p_\perp^2\to  p^-={p_\perp^2+m^2\over p^+},$
 a relativistic kinetic energy which does not contain
a square root operator. This
allows the  separation of  center of mass
and relative coordinates, so that the computed wave functions are frame
independent.

The use of the light front is particularly relevant for calculating form
factors, which are probability amplitudes for an nucleon to absorb a four
momentum $q$ and remain  a nucleon. The initial and final nucleons 
have different total momenta. This means that the final nucleon is a
boosted nucleon, with 
 different wave function than the initial nucleon.  In general, performing
 the boost is difficult for large values of $Q^2=-q^2$. However the light
 front technique allows one to set up the calculation so that the boosts are
 independent of interactions. Indeed, the wave functions are functions of
 relative variables and are independent of frame.

\subsection            {Definitions}

\newcommand{\boldsigma}{\mbox{\boldmath $\sigma$}}

Let us define the basic quantities concerning us here. These are the
independent form factors defined by
\begin{equation}
\left< N,\lambda ' p' \left| J^\mu \right| N,\lambda p\right> =
\bar u_{\lambda '}(p') \left[ F_1(Q^2)\gamma^\mu + {\kappa F_2(Q^2) \over
2 M_N}i\sigma^{\mu\nu}(p'-p)_\nu \right] u_\lambda (p).
\ee
The 
Sachs form factors are defined by the equations:
\bea
G_E = F_1 - {Q^2 \over 4M_N^2}\kappa F_2,\; 
G_M = F_1 +  \kappa F_2\label{sachsdefs}.\eea

There is an alternate 
  light front interpretation, based on field theory, in which one uses
the  ``good" component  of the current, $J^+,$ to 
 suppress  the effects of quark-pair terms. Then, using nucleon light-cone
 spinors: 
\begin{eqnarray}
F_1(Q^2) ={1 \over 2P^+} N,\langle\uparrow\left| J^+\right| N,
\uparrow\rangle, 
Q\kappa  F_2(Q^2) ={-2M_N \over 2P^+}\langle N,\uparrow\left|
J^+\right| N,\downarrow\rangle.
\end{eqnarray}


\section{Why I am Giving This Talk}

In 1996, Frank, Jennings \& I \cite{Frank:1995pv} examined 
the point-like-configuration    idea of
Frankfurt \& Strikman\cite{Frankfurt:cv}. We needed to start with a
relativistic model of the free nucleon. The resulting form factors
are shown in Figs.~10 and 11 of our early paper. 
The function $G_M$ was constrained\cite{Schlumpf:ce}
by experimental data to define the parameters of the model, but we
predicted  a very strong decrease of $G_E/G_M$ as a function of $Q^2$. This
decrease has now been measured as a real effect, but
the task of explaining its meaning  
 remained  relevant.
That was the purpose of our second paper\cite{Miller:2002qb} in which 
 imposing Poincar\'{e} invariance was shown to lead
 to substantial violation of the
helicity conservation rule as well as an analytic result that the ratio
$QF_2/F_1$ is constant for the $Q^2$ range of the Jefferson Laboratory
experiments. Although the second paper
is new, the model is the same. 
Ralston {\em et al.}\cite{ralston} have been talking about
non-conservation of helicity for a long time.

\section {  Three-Body Variables and Boost}
We use 
light front coordinates for the momentum 
of each of the $i$ quarks, such that
${\bf p}_i =
(p^+_i,{\bf p}_{i\perp}), p^-=(p_\perp^2+m^2)/p^+.$ The  total
(perp)-momentum
is 
$\bf {P}= {\bf p}_1+{\bf p}_2+ {\bf p}_3,$ the plus components of the
momenta are denoted as 
\be \xi={p_1^+\over p_1^++p_2^+}\;,
\qquad
\eta={p_1^++p_2^+\over P^+},\ee
and the perpendicular relative coordinates are given by
\be {\bf k}_\perp =(1-\xi){\bf p}_{1\perp}-\xi {\bf p}_{2\perp}\;, \quad
{\bf K}_\perp =(1-\eta)({\bf p}_{1\perp}
+{\bf p}_{2\perp})-\eta {\bf p}_{3\perp}.\ee
In 
the  center of mass  frame we find:
\be
{\bf p}_{1\perp}={\bf k}_\perp+\xi {\bf K}_\perp,\;\;\quad
{\bf p}_{2\perp}=-{\bf k}_\perp+(1-\xi){\bf K}_\perp\;,\quad 
{\bf p}_{3\perp}=-{\bf K}_\perp .\ee
The coordinates $\xi,\eta, \bfk,\bfK$ are
all relative coordinates so that one obtains a 
 frame independent wave function
   $\Psi({\bf k}_\perp,{\bf K}_\perp,\xi,\eta).$

Now consider the computation of a form factor, taking quark 3 to be the one
struck by the photon. One works in a special set of frames with $q^+=0$ and
$Q^2=\bfq_\perp^2$,
so that the value of $1-\eta$ is not changed by the photon. The
coordinate $\bfp_{3\perp}$ is changed to $\bfp_{3\perp}+\bfq_\perp,$
so  only one relative momentum, $\bfK_\perp$ is changed:
\bea{\bfK'}_\perp =(1-\eta)({\bf p}_{1\perp}
+{\bf p}_{2\perp})-\eta ({\bf p}_{3\perp}+{\bf q}_\perp)\;
=\bfK_\perp-\eta\bfq_{\perp},\quad
\bfk'_\perp=\bfk_\perp,\qquad  \eea

The arguments of the spatial wave function are
taken as the mass-squared operator for a non-interacting system:
\be M_0^2\equiv\sum_{i=1,3} p^-_i\;P^+-P_\perp^2=
{K_\perp^2\over \eta(1-\eta)}+
{k_\perp^2+m^2\over \eta\xi(1-\xi)} +{m^2\over 1-\eta}. \ee
This  is a relativistic version of the  square of a the
relative three-momentum. Note that      the absorption of a photon 
changes the  value to:
\be{{M_0}'}^2=
{(K_\perp-\eta q_\perp)^2\over \eta(1-\eta)}+
{k_\perp^2+m^2\over \eta\xi(1-\xi)} +{m^2\over 1-\eta}.\ee

\section { Wave function}
Our wave function is based on symmetries. The wave function is anti-symmetric,
a function of relative momenta, independent of reference frame, an eigenstate
of the spin operator and rotationally invariant (in a specific well-defined
sense). The use of symmetries is manifested in the construction of such 
wave functions, as originally described by
 Terent'ev \cite{bere76},               Coester\cite{chun91} and their
 collaborators. 
A schematic  form of the wave functions is
 \be
\Psi(p_i)=\Phi(M_0^2)
u(p_1) u(p_2) u(p_3)\psi(p_1,p_2,p_3),\quad p_i=\bfp_i  s_i,\tau_i\ee
where $\psi$ is a spin-isospin color amplitude factor,
the $p_i$ are expressed in terms of relative coordinates,  the
$u(p_i)$ are ordinary 
  Dirac spinors and $\Phi$ is a spatial wave function.

We take the the spatial wave function from  Schlumpf\cite{Schlumpf:ce}: \bea
\Phi(M_0)={N\over (M^2_0+\beta^2)^{\gamma}}\;,  
\beta =0.607\;{\rm GeV}, \; \gamma=3.5,\; m = 0.267\; {\rm GeV}.
\label{params}\eea
 The value
of 
$\gamma$ is chosen that $ Q^4G_M(Q^2) $   is approximately  constant for
$Q^2>4\; {\rm GeV}^2$ in accord with experimental data. The parameter 
$\beta$ helps govern the values of the perp-momenta allowed by the
wave function $\Phi$ and is closely related to the  rms charge radius,
and $m$ is mainly determined by the magnetic moment of the proton.

At this point the wave function and the calculation  are 
completely defined. One could evaluate
the form factors as $\langle \Psi\vert J^+\vert \Psi\rangle$ and obtain the
previous numerical results\cite{Frank:1995pv}. 

\section { Simplify  Calculation- Light Cone Spinors}

The operator $J^+\sim \gamma^+$ acts its
  evaluation is simplified by using light cone
spinors. These      solutions of the free Dirac equation, related to ordinary
Dirac spinors by a unitary transformation,   conveniently satisfy:
\bea \bar u_L(p^+,\bfp',\lambda')\gamma^+u_L(p^+,\bfp ,\lambda)=
2\delta_{\lambda\lambda'}p^+. \eea 
To take advantage of this,
re-express the wave function in terms of light-front spinors using
the  completeness relation: 
$1= \sum_\lambda u_L(p,\lambda)\bar u_L(p,\lambda).$ We then find
  \bea
&&\Psi(p_i)=u_L(p_1,\lambda_1) u_L(p_2,\lambda_2) u_L(p_3,\lambda_3)
\psi_L(p_i,\lambda_i),\\
&&\psi_L(p_i,\lambda_i)\equiv
[\bar u_L(\bfp_1,\lambda_1)u(\bfp_1, s_1)]
[\bar u_L(\bfp_2,\lambda_2)u(\bfp_2, s_2)]\nonumber\\ \times
&&[\bar u_L(\bfp_3,\lambda_3)u(\bfp_3, s_3)]\;
\psi(p_1,p_2,p_3).\eea This is
the  very same  $\Psi$ as before, it is just that now it is
easy  to compute the matrix elements of the $\gamma^+$ operator. 

The  unitary transformation is  also known as the Melosh rotation.
The basic point is that one may evaluate the coefficients in terms of Pauli
spinors: $\vert \lambda_i\rangle,\vert s_i\rangle,$
with $\langle \lambda_i\vert { R}_M^\dagger(\bfp_i)\vert s_i\rangle
\equiv \bar u_L(\bfp_i,\lambda_i)u(\bfp_i, s_i)$. It is easy to show that
  \be \langle \lambda_3\vert 
{ R}_M^\dagger(\bfp_3)\vert s _3\rangle= \langle \lambda_3\vert
\left[ {m+(1-\eta)M_0+i{\boldsigma}\cdot({\bf n}\times {\bf p}_3)\over
\sqrt{(m+(1-\eta)M_0)^2+p_{3\perp}^2}}\right]\vert s_3\rangle.
\label{melosh}\ee
The important effect resides in the term $({\bf n}\times {\bf p}_3)$ which
originates from the lower components of the Dirac spinors. This  large
  relativistic spin effect can be summarized:
  the effects of relativity are to replace 
 Pauli spinors by     Melosh rotation operators acting on Pauli spinors.
 Thus 
\be \vert\uparrow \bfp_i\rangle\equiv{ R}_M^\dagger(\bfp_i)
\pmatrix{1\cr 0},\;\vert\up\bfp_3\rangle\ne \vert\up\rangle.\ee
\section{  Proton $F_1,F_2$-Analytic Insight}

The analytic insight is based on Eq.~(\ref{melosh}). Consider
high momentum transfer such that
$Q=\sqrt{\bfq_\perp^2}\gg \beta=560$ MeV.  {\em Each} of the quantities: 
$M_0\;,M_0'\;,\bfp_{3\perp},\;\bfp_{3\perp}$ can be of order $q_\perp,$ 
so   the spin-flip term is as large as the non-spin flip term. In particular,
($s_3=+1/2$)     
 may correspond to ($\lambda_3=-1/2)$, so the spin of the  
  struck quark $\ne$ proton  spin.
This means that
there is no hadron helicity selection rule\cite{ralston,Braun:2001tj}.

The effects of the 
  lower components of  Dirac spinors, which cause the
  spin flip term $\boldsigma\times \bfp_3$,
  are the same as having a  non-zero $L_z$,
  if the wave functions are expressed in the light-front basis.
  See Sect.~9. 

We may now  qualitatively understand the numerical results, since 
\begin{eqnarray}
F_1(Q^2) &=& 
\int\! {d^2\!q_\perp d\xi\over \xi(1-\xi)}{ d^2K_\perp d\eta\over\eta(1-\eta)}\;
\cdots
\;\langle\up\bfp_3'\vert
\up(\bfp_3)\rangle \\
Q\kappa F_2(Q^2) &=& 2M_N
\int\! {d^2\!q_\perp d\xi\over \xi(1-\xi)} 
{d^2K_\perp d\eta\over\eta(1-\eta)}\;
\cdots
\langle\up\bfp_3'\vert
\down(\bfp_3)\rangle, 
\end{eqnarray}
where the $\cdots$ represents common factors. The term
$F_1\sim\langle\up\bfp_3'\vert\up\bfp_3\rangle$ is a spin-non-flip
term and 
$F_2\sim\langle\up\bfp_3'\vert\up\bfp_3\rangle$ depends on the spin-flip term.
In doing the integral each of the momenta, and $M_0,M_0'$ can take the
large value $Q$ for some regions of the integration. Thus in the integral
\bea
\langle\up\bfp_3'\vert\up\bfp_3\rangle
\sim {Q\over Q},\;\quad
\langle\up\bfp_3'\vert\down\bfp_3\rangle\sim {Q\over Q},\eea
so that 
 $F_1$ and $ QF_2$ have the same $Q^2$ dependence. This
is shown in Fig.~1.
\begin{figure}
\unitlength1cm 
\begin{picture}(10,8)(0,-8.)
  \includegraphics{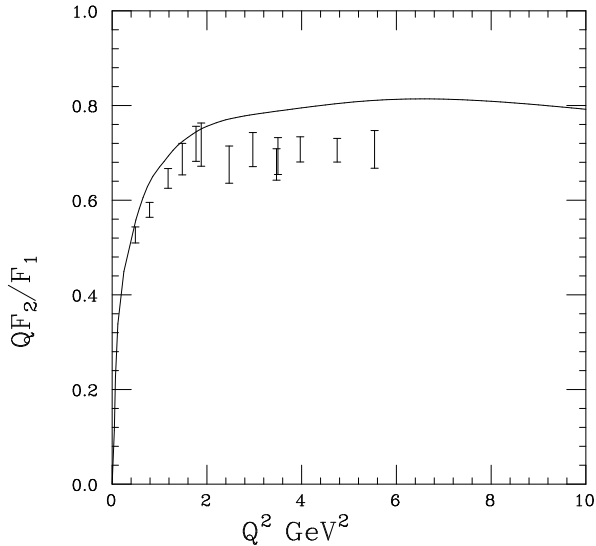}
\end{picture}
\label{fig:1}
\caption{Calculation of Refs.~$^{2,5}$, 
  data 
  are from  Jones {\em et al.}$^{10}$ 
and from  Gayou {\em et al.}.} 
\end{figure}

\section{Relation between ordinary Dirac Spinors and $L_z$}

Our use of ordinary Dirac spinors corresponds
to the use of a non-zero $L_z$ in the light front basis. 
We may represent Dirac spinors as Melosh rotated Pauli spinors, and
this is sufficient to show $L_z\ne0$.

It is worthwhile to
 consider the pion as an explicit example. Then our version of 
 the  light-front wave function $\chi_\pi$ would be\cite{cc}:
\be
\chi_\pi(k^+,\bfk_\perp,\lambda,\lambda')\propto
\langle\lambda\vert i\sigma_2 m-(k_1-i\sigma_3k_2)\vert\lambda'\rangle,\ee
 while the Gousset-Pire-Ralston\cite{ralston}
 pion
Bethe-Salpeter  amplitude $\Phi$ is 
   \be \Phi=P_{0\pi} \notp_\pi+P_{1\pi}[\notp_\pi,\notk_\perp],\ee
     where $p_\pi$ is the pion total momentum,  $P_{i\pi}$ are scalar
     functions of relative momentum, and the term with $P_{1\pi}$
     is the one which carries orbital angular momentum.
     The relation\cite{ls} between the Bethe-Salpeter
     amplitude and the light-front wave function $\phi_\pi$ is
     \be \phi_\pi(k^+,\bfk_\perp,\lambda,\lambda')=
     \bar{u}_L(k^+,\bfk_\perp,\lambda)\gamma^+\Phi\gamma^+
     v_L(P_\pi^+-k^+,-\bfk_\perp,\lambda').\ee
     Doing  the Dirac algebra and choosing  suitable
functions      $P_{i\pi}$, leads to 
$ \chi_\pi=\phi_\pi.$
The Melosh transformed Pauli spinors, which account for the lower components of
the ordinary Dirac spinors, contain the non-zero angular momentum of the
 wave function $\Phi$.

\section{Neutron Charge Form Factor}
The neutron has no charge, $ G_{En}(Q^2=0)=0$, and the square of its charge
radius is determined from the low $Q^2$ limit as
$G_{En}(Q^2)\to-Q^2R^2/6.$ The quantity $R^2$ is well-measured\cite{nrm} as
$R^2=-0.113 \pm 0.005 \;$fm$^2$. The Galster 
parameterization\cite{galster}
 has been used
to represent the data for $Q^2<0.7\; {\rm GeV}^2$.

Our proton respects charge symmetry, the interchange of $u$ and $d$ quarks,
so  it  contains a prediction for neutron form factors.
This is shown in Fig.~2.
\begin{figure}
\unitlength1cm
\begin{picture}(10,8)(0,-8)
\includegraphics{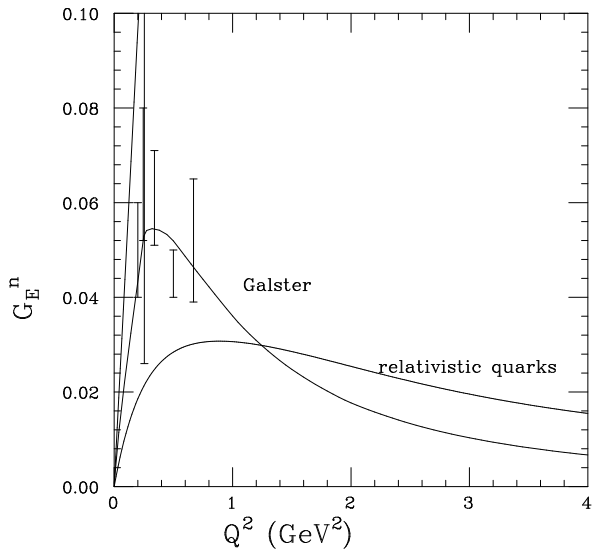}
\end{picture}
\label{fig:2}
\caption{Calculation of $G_E^n.$
  The data
  are from  Ref.~$^{15}$, with more expected soon$^{16}$. }
\end{figure}
The resulting curve labeled relativistic quarks is both large and small. It is
very small at low values of $Q^2$. Its slope at $Q^2=0$ is too small by a factor
of five, if one compares with the straight line. But at larger values of $Q^2$
the prediction is relatively large.

Our model gives $R^2_{\rm model}=-0.025$ fm$^2$, about five times
smaller than the data. The small value can be understood in terms of $F_{1,2}$.
\begin{figure}
\unitlength1cm
\begin{picture}(10,8)(0,-8.5)
\includegraphics{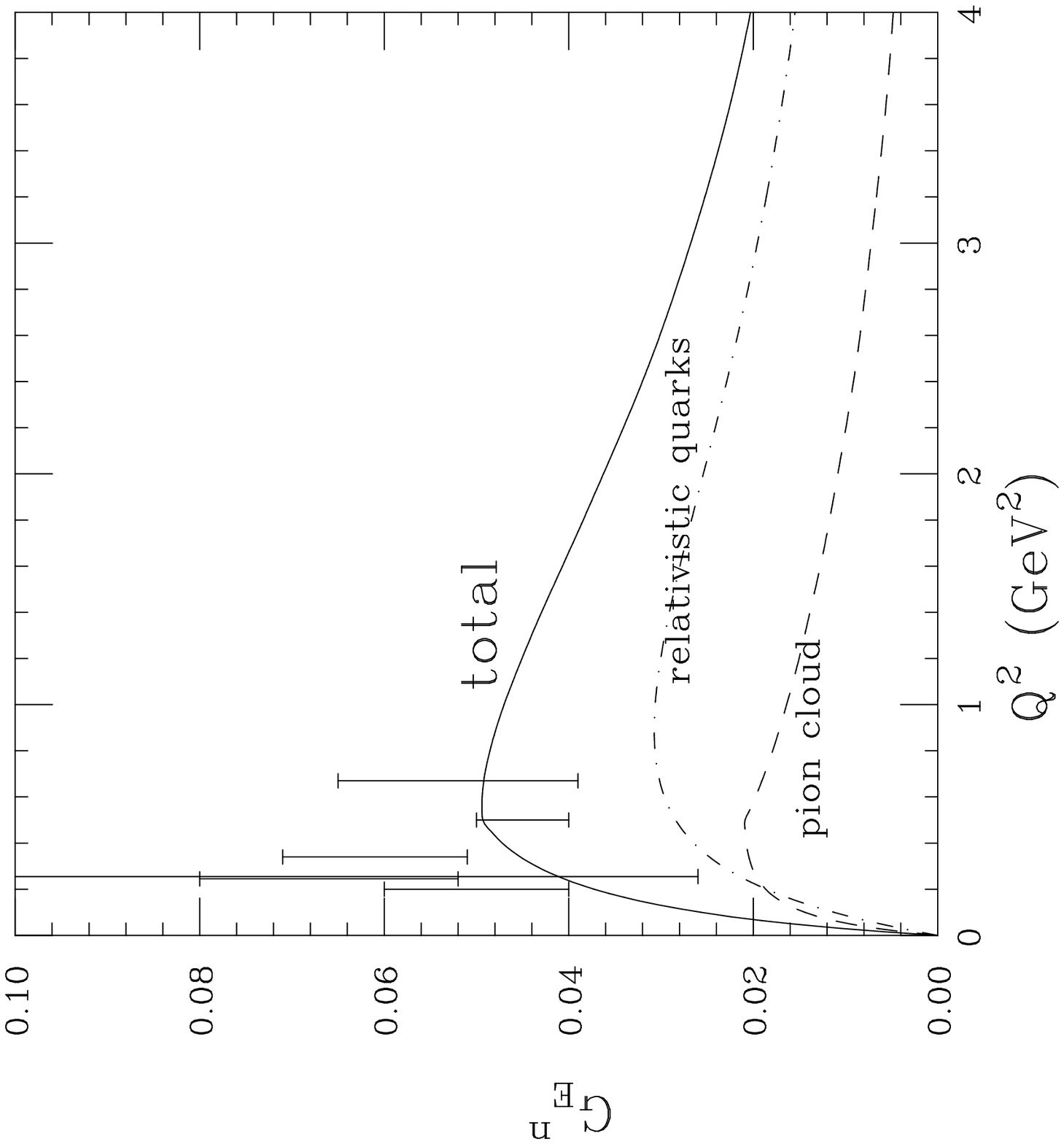}
\end{picture}
\label{fig:ratio}
\caption{Light front cloudy bag model LFCBM Calculation of $G_E^n$.}
\end{figure}
Taking the  definition (\ref{sachsdefs}) for small values of $Q^2$
gives
\bea-Q^2R^2/6=-Q^2R_1^2/6 -\kappa_n Q^2/4M^2=-Q^2R_1^2/6 -Q^2R_F^2/6,\eea
where the Foldy contribution, 
$R_F^2=6 \kappa_n/4M^2=-0.111\; {\rm fm}^2$, is  in good agreement with
the experimental data. That  a point
particle with a magnetic moment can explain the charge radius has led 
some  to state that $G_E$ is  not a measure
of  the structure of the neutron. However,
one must include the $Q^2$ dependence of $F_1$ which gives $R_1^2$. In our
model $R_1^2=+0.086\; {\rm fm}^2$ which nearly cancels the effects of $R_F^2$.
Isgur\cite{Isgur:1998er} showed that
this cancellation is a natural consequence of including the relativistic
effects of the lower components of the Dirac spinors. Thus our relativistic
effects are standard. We need another source of $R^2$. This is  the
pion cloud.
\section{Pion Cloud  and the Light Front Cloudy Bag Model}
The effects of chiral symmetry require that
sometimes a physical nucleon can be a  bare nucleon emersed in   a pion
cloud. An incident photon can interact electromagnetically with
a bar nucleon, a pion in flight or with a nucleon while a pion is present.
These effects were included in the cloudy bag model, and are especially
pronounced for the neutron. Sometimes the neutron can be a proton plus a
negatively charged pion. The tail of the pion distribution extends far out into
space (see Figs. 10 and 11) of Ref.\cite{cbm}, so that the square of the charge
radius is negative.

It is necessary to modernize the cloudy bag model, so as to make it
relativistic. This involves using photon-nucleon form factors from our model,
using a relativistic $\pi$-nucleon form factor, and treating the pionic
contributions relativistically by doing  a light front calculation. This has
been done. The result is the light front cloudy bag model, 
 and the preliminary results are shown in Fig.~3. We see that the pion
cloud effects are important for small values of $Q^2$ and, when combined with
those of the relativistic quarks coming from the bare nucleon, leads to a
good description of the low $Q^2$ data. The total value of $G_E$ is substantial
for large values of $Q^2$.
\section{Summary}
\poinc invariance is needed to describe the new exciting experimental results.
Ordinary Dirac spinors carry light front orbital angular momentum. Including
the effects of these spinors, in a way such that the proton is an eigenstate of
spin leads naturally to the result that $QF_2/F_1$ is constant
for values of $Q^2$ between 
2 and about 20 GeV$^2$.

The prediction of hadron helicity conservation is that $Q^2F_2/F_1$ is constant,
so we see that this is not respected in present data and there is no
need to expect it to hold for a variety of exclusive reactions occurring
at high $Q^2\le 5.5$ GeV$^2$.
Examples include the anomalies seen in $pp$ elastic scattering
and the large spin effects seen in the reactions $\gamma d\to np$ and
$\gamma p\to \pi^0 p$.

The results for the neutron $G_E$ can be concisely stated. At small values of
$Q^2$ the effects of a pion cloud is needed to counteract the relativistic
effects which cancel the effects of the Foldy term. At large values of
$Q^2$ relativistic effects give a ``large'' value of $G_E$; large in the sense
that this form factor is predicted to be larger than that of the Galster
parameterization.

At the time of this workshop, I had not yet used the light front
cloudy bag model to compute proton form factors or the neutron's $G_M$.
Including the
effects of the pion cloud (with a    parameter  to describe
the pion-nucleon form factor) allows the use  of
different quark-model parameters. 
The result is an excellent description of all
four nucleon electromagnetic form factors, and I  plan to publish that soon.

\section*{Acknowledgments} This work is partially supported by the U.S. DOE.
I thank R.~Madey for encouraging me to compute the neutron form factors.


\begin{thebibliography}{99}

\bibitem{Frank:1995pv}
M.R.~Frank, B.K.~Jennings and G.A.~Miller,
Phys.\ Rev.\ C {\bf 54}, 920 (1996).
\bibitem{Frankfurt:cv}
L.~L.~Frankfurt and M.~I.~Strikman,
Nucl.\ Phys.\ B {\bf 250}, 143 (1985).




\bibitem{Schlumpf:ce}
F.~Schlumpf,
U. Zurich Ph. D. Thesis, hep-ph/9211255.



\bibitem{Miller:2002qb}
G.~A.~Miller and M.~R.~Frank,
nucl-th/0201021 to appear Phys.\ Rev.\ C.
\bibitem{ralston} 
P.~Jain, B.~Pire and J.~P.~Ralston,
Phys.\ Rept.\  {\bf 271}, 67 (1996)
;
T.~Gousset, B.~Pire and J.~P.~Ralston,
Phys.\ Rev.\ D {\bf 53}, 1202 (1996)
\bibitem{bere76}
V.~B. Berestetskii and M.~V. Terent'ev.
\newblock Sov. J. Nucl. Phys. {\bf 25}, 347 (1977).
\bibitem{chun91}
P.~L. Chung and F.~Coester.
\newblock {\em Phys. Rev. D {\bf 44}},  229, (1991).
\bibitem{Braun:2001tj}
V.~M.~Braun, A.~Lenz, N.~Mahnke and E.~Stein,
Phys.\ Rev.\ D {\bf 65}, 074011 (2002).

\bibitem{Jones:1999rz}
M.~K.~Jones {\it et al.}, 
Phys.\ Rev.\ Lett.\  {\bf 84}, 1398 (2000)

\bibitem{Gayou:2001qd}
O.~Gayou {\it et al.}, 
Phys.\ Rev.\ Lett.\  {\bf 88}, 092301 (2002).
\bibitem{cc} P.L. Chung, F. Coester and W.N. Polyzou, Phys. Lett 205B,545 
(1988).
\bibitem{ls}
H.~H.~Liu and D.~E.~Soper,
Phys.\ Rev.\ D {\bf 48}, 1841 (1993).
\bibitem {nrm}
S. Kopecky {\it et al.},
Phys. Rev. Lett.  {\bf74}, 2427 (1995)

\bibitem{galster}
    S.~Galster{\it et al.},
Nucl.\ Phys.\ B {\bf 32}, 221 (1971).
\bibitem{nrefs}
T.~Eden {\it et al.},
Phys.\ Rev.\ C {\bf 50}, 1749 (1994);
M.~Meyerhoff {\it et al.},
Phys.\ Lett.\ B {\bf 327}, 201 (1994);
M.~Ostrick {\it et al.},
Phys.\ Rev.\ Lett.\  {\bf 83}, 276 (1999).
J.~Becker {\it et al.},
Eur.\ Phys.\ J.\ A {\bf 6}, 329 (1999).
I.~Passchier {\it et al.},
Phys.\ Rev.\ Lett.\  {\bf 82}, 4988 (1999)
D.~Rohe {\it et al.},
Phys.\ Rev.\ Lett.\  {\bf 83}, 4257 (1999).
H.~Zhu {\it et al.}  
Phys.\ Rev.\ Lett.\  {\bf 87}, 081801 (2001)


\bibitem{madey}Jefferson Laboratory Experiment 93-038,
  R. Madey Spokesperson; R.Madey, for the Jlab E93-038 collaboration,
``Neutron Electric Form Factor Via Recoil Polarimetry'',
contribution to Baryons 2002.

\bibitem{Isgur:1998er}
N.~Isgur,
Phys.\ Rev.\ Lett.\  {\bf 83}, 272 (1999)
\bibitem{cbm} S.\ Th\'eberge, A.\ W.\ Thomas and G.\ A.\ Miller,
Phys. Rev. {\bf D22} (1980) 2838;  (1981) 2106, A.\ W.\ Thomas,
S.\ Th\'eberge, and G.\ A.\ Miller, Phys. Rev. {\bf D24} (1981) 216; 
S.~Th\'eberge, G.~A.~Miller and A.~W.~Thomas,
Can.\ J.\ Phys.\  {\bf 60}, 59 (1982).
G.~A.~Miller, A.~W.~Thomas and S.~Th\'eberge,
Phys.\ Lett.\ B {\bf 91}, 192 (1980).


                               
\end{thebibliography}
\end{document}